\documentclass[12pt]{iopart}
\usepackage{graphicx}
\usepackage{iopams}  
\usepackage{harvard}

\expandafter\let\csname equation*\endcsname\relax
\expandafter\let\csname endequation*\endcsname\relax
\usepackage{amsmath}
\usepackage{cite}

\usepackage{xspace}
\usepackage[usenames,dvipsnames]{color}
\usepackage{soul}
\usepackage{epstopdf}

\newcommand{\eryso}[0]{Er$^{3+}$:Y$_2$SiO$_5$\xspace}

\newcommand{\yso}[0]{Y$_2$SiO$_5$\xspace}

\newcommand{\dun}[0]{ {\bf D$_1$}\xspace}
\newcommand{\ddeux}[0]{ {\bf D$_2$}\xspace}
\newcommand{\baxis}[0]{ {\bf b}\xspace}

\begin{document}

\title[Optical memory  bandwidth and multiplexing capacity in the erbium window]{Optical memory bandwidth and multiplexing capacity in the erbium telecommunication window}

\author{J. Dajczgewand$^1$, R. Ahlefeldt$^1$, T. B\"ottger$^2$, A. Louchet-Chauvet$^1$, J.-L. Le Gou\"et$^1$, T. Chaneli\`ere$^1$}
\address{$^1$ Laboratoire Aim\'e Cotton, CNRS-UPR 3321, Univ. Paris-Sud, B\^at. 505, 91405 Orsay cedex, France}

\address{$^2$ Department of Physics \& Astronomy, 2130 Fulton Street, University of San Francisco, San Francisco, CA 94117, USA}

\ead{thierry.chaneliere@u-psud.fr}

\begin{abstract}
We study the bandwidth and multiplexing capacity of an erbium-doped optical memory for quantum storage purposes. We concentrate on the protocol  ROSE (Revival of a Silenced Echo) because it has the largest potential multiplexing capacity. Our analysis is applicable to other protocols that involve strong optical excitation. We show that the memory performance is limited by instantaneous spectral diffusion and we describe how this effect can be minimised to achieve optimal performance.
\end{abstract}
%Uncomment for PACS numbers title message§
%32.10.Dk	Electric and magnetic moments, polarizabilities
%42.50.Ex	Optical implementations of quantum information processing and transfer
%42.50.Md	Optical transient phenomena: quantum beats, photon echo, free-induction decay, dephasings and revivals, optical nutation, and self-induced transparency
%78.47.jf	Photon echoes
\submitto{\NJP}
\maketitle
%\tableofcontents

\section{Introduction}
A quantum memory is characterized by different figures of merit. The efficiency and the storage time have long been understood to be paramount, but for the practical application of quantum repeaters for which memories are the crucial element  \cite{Sangouard11}, the bandwidth and the multimode capacity also need to be considered. The bandwidth is important because it scales the overall transmission rate of the quantum repeater, while the multimode capacity determines the number of qubits that can be stored in parallel.

The ability to multiplex a quantum repeater is particularly important because it significantly boosts the global  performance \cite{Collins07}. Multiplexed memories \cite{Simon07, Simon10} are mainly of two types: spatial and spectral. In the former, the atomic medium is divided into spatially independent sub-ensembles that each store an information bit. The realizations of such spatially multiplexed memories have involved, for example, laser cooled atomic clouds \cite{Lan09, Golubeva11}. Temporal multiplexing, which makes use of the spectral dimension, is available in media that exhibit large inhomogeneous broadening \cite{Nunn08}.  Studies of temporal multiplexing have concentrated on doped solids, as high ratio of inhomogeneous to homogeneous linewidths allows a large spectral capacity. The storage of a train of pulses has been demonstrated, for example, using the Atomic Frequency Comb protocol (AFC)  \cite{Afzelius09,Bonarota11}. The latter  protocol has been also exploited very recently to show direct spectral multiplexing \cite{sinclair14}, further expanding the possibilities of the protocol.

The potential memory bandwidth in an inhomogeneous sample is given by the absorption profile, but the same inhomogeneities that allow a large bandwidth also induce dipole dephasing. For this reason, the storage protocols used in inhomogeneous samples are related to the photon-echo technique \cite[and references therein]{Tittel10,Lvovsky09}, which  compensates the dephasing while maintaining the large bandwidth offered by an inhomogeneous sample. The two-pulse photon echo technique itself has limitations that render it unsuitable for practical storage \cite{Ruggiero09, Sangouard10}, and the various storage protocols that have been proposed avoid these problems in different ways. The Controlled Reversible Inhomogeneous Broadening protocol (CRIB) \cite{Nilsson05, Moiseev01, Alexander06, Kraus06} replaces the natural broadening with an artificial narrow line that can be controlled by electric fields, while the AFC requires a series of evenly spaced narrow peaks \cite{Afzelius09}. Both protocols require an initial optical pumping sequence to carve the  necessary absorption profile. CRIB and AFC have different multiplexing capacities \cite{Nunn08} essentially scaling as the number of narrow peaks left after the preparation.

We recently proposed the protocol Revival Of a Silenced Echo (ROSE) \cite{Damon11}. Because this protocol does not require any state preparation, it preserves the complete inhomogeneous line. This is a feature shared with the protocol HYPER \cite{Hyper}, which belongs to the same family of protocols. The time-to-bandwidth-product is not intrinsically limited by the scheme itself but by technical parameters such as the inhomogeneous linewidth, coherence lifetime and available laser power.

We implement ROSE in an erbium doped sample compatible with the telecommunication C-band. Previously, we reported a large efficiency in this range \cite{Dajczgewand14}. In the present paper we focus on the bandwidth and evaluate the multimode capacity of the sample. We identify instantaneous spectral diffusion (ISD) as a limiting factor.

The paper is organized as follows. We first restate the basics of the protocol, describing the use of chirped pulses as adiabatic passages, an originality of the protocol. These pulses are used to improve the inversion of  the atoms in the storage bandwidth, but they also introduce a dependancy between the storage time and bandwidth, which will be explained in detail.  We then study the storage bandwidth in the range $\sim$ 1 MHz -  10 MHz. The observed behavior is explained by the influence of ISD, which we investigate in the subsequent section.  Finally, we use  our results to determine the optimal performance of the memory in the \eryso sample under study.

\section{ROSE efficiency and bandwidth scaling}\label{section:scaling}
\begin{figure}[h]
\centering{\includegraphics[width=0.6\paperwidth]{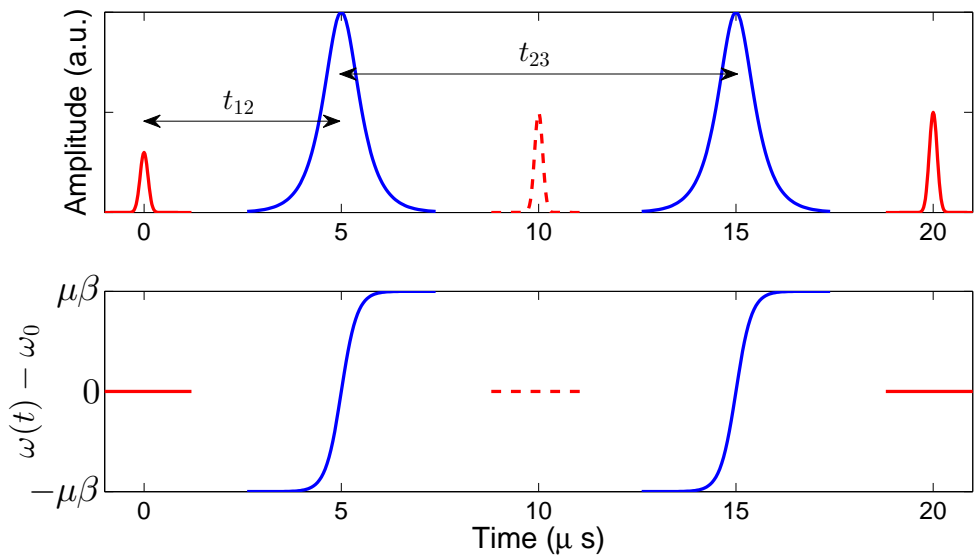}}
\caption{ROSE temporal scheme. The delay between the signal and the first CHS rephasing pulse is $t_{12}$ and between the first and the second CHS is $t_{23}$. In this example, $t_{23}=2t_{12}=10 \mu s$. Top: the pulse sequence. The incoming signal (at $t=0$ in red) is absorbed by the medium. The echo (at $t=2 t_{23}$ in red) is retrieved after the application of two rephasing pulses (in blue). We also represent the silenced echo (at t=$2 t_{12}$ in dashed red) corresponding to the silenced rephasing. The silencing of this first echo is achieved by spatial phase mismatching; the two $\pi$-pulses are sent along a different wave vector to the signal pulse.  Bottom: The instantaneous frequencies of the pulses $\omega \left( t \right)$. The signal is a monochromatic pulse whose bandwidth is Fourier transform limited. The rephasing pulses are chirped, taking the form of a Complex Hyperbolic Secant (see text for details).}
\label{fig:Rose_scheme}
\end{figure}

ROSE is a descendant of the two-pulse photon echo. It employs two rephasing pulses instead of one to avoid the inversion of the medium and the associated spontaneous emission noise \cite{Ruggiero09}. Short $\pi$-pulses can be advantageously replaced by chirped pulses that perform a rapid adiabatic passage \cite{de_Seze05, Pascual-Winter13}. Specifically, we use Complex Hyperbolic Secant (CHS) pulses  \cite{Damon11} in the typical time sequence presented in fig.\ref{fig:Rose_scheme}.

The CHS pulse envelope is a hyperbolic secant $\Omega(t)$ associated with a hyperbolic tangent frequency sweep $\omega \left( t \right)=\displaystyle \frac{d \phi(t) }{dt}$ ($\phi(t)$ is the phase) about the central frequency $\omega_0$:
\begin{equation} \begin{array}{ll}\label{CHS} \Omega \left( t \right) = \Omega_0 \mathrm{sech}\left(\beta t\right) \\[0.4cm] \omega \left( t \right) = \omega_0 + \mu \beta \mathrm{tanh}\left(\beta t\right) \end{array} \end{equation}
where $1/\beta$ is the pulse duration and $\mu$ a constant. This can be written in a more compact manner by defining the complex envelope 
 $\displaystyle \Omega \left( t \right)e^{i\phi(t)}$ as
 \begin{equation}
 \Omega \left( t \right)e^{i\phi(t)} = \Omega_0 \left[\mathrm{sech}\left(\beta t\right)\right]^{1-i\mu}
 \end{equation}
  justifying the term Complex Hyperbolic Secant.

To ensure that the CHS pulse coherently drives the Bloch vector, the adiabatic condition must be satisfied \cite{de_Seze05, Pascual-Winter13}:
\begin{equation}\label{adiabcond} \mu \beta^2  \ll  \Omega_0^2 \end{equation}
When this condition is satisfied, {the atoms are uniformly inverted over the bandwidth $\mathcal{B}=2 \mu \beta$ covering the interval $[-\mu \beta, \mu \beta]$ when $\mu \geq 1$ } \cite{Hioe84, Silver85}. Because the adiabatic condition (eq. \ref{adiabcond}) puts an upper bound on the chirp rate $ \mu \beta^2$ for a given laser power $ \Omega_0^2$, the CHS duration $1/\beta$ and the inversion bandwidth $2 \mu \beta$  are not fully independent. This general statement should be kept in mind when CHS pulses are used for quantum storage as in ROSE. 

This dependence can be more clearly seen by rewriting the adiabatic condition eq.\eqref{adiabcond} as $\mathcal{B} \times \beta  \ll 2 \Omega_0^2$. This means that as the protocol bandwidth is increased, longer pulses must be used. As a consequence (see in fig.\ref{fig:Rose_scheme}), the temporal scale has to be expanded accordingly to prevent the overlap of the CHS pulses with the signal and echo or with each other.

Through experiments, we have verified that the adiabatic condition is well satisfied by  keeping \cite{de_Seze05}:
\begin{equation}\label{adiabcond_exp} \mu \beta^2 = \frac{ \Omega_0^2}{4} \Longleftrightarrow  \mathcal{B} \times \beta = \frac{ \Omega_0^2}{2} \end{equation}
As previously mentioned, to prevent the overlap of the signal and echo with the CHS pulses and between the CHS pulses themselves, we keep (see fig.\ref{fig:Rose_scheme} for the definitions):
\begin{equation}  t_{23}^\mathrm{min}=2t_{12}^\mathrm{min}=\frac{8 \pi}{\beta} \label{t23_beta}
\end{equation}
As can be seen in fig.\ref{fig:Rose_scheme} the minimum delay between any two pulses  is then $t_{12}^\mathrm{min}=\displaystyle \frac{4 \pi}{\beta}$. The two constraints (\ref{adiabcond_exp}) and (\ref{t23_beta}) correspond to appropriate practical conditions for implementing ROSE.

The total storage time between the signal and the ROSE echo is given by $2 t_{23}$. As a consequence, the efficiency $\eta$ is given by \cite{Damon11}:
\begin{equation}
\eta=(\alpha L)^{2} \exp({-\alpha L}) \exp\left({-\frac{4 t_{23}}{T_2}}\right) = \eta_0 \exp\left({-\frac{4 t_{23}}{T_2}}\right)
\label{eq:eff}
\end{equation}
where {$\eta_0=(\alpha L)^{2} \exp({-\alpha L})$ is constant for a given optical depth. We analysed the dependency on $\alpha L$ in } \cite{Dajczgewand14}.

This expression can be reformulated to be a function of the bandwidth rather than the storage time as the two are related by equations \eqref{adiabcond_exp} and \eqref{t23_beta} when the minimum delays  $t_{23}^\mathrm{min}$ and $t_{12}^\mathrm{min}$ are maintained:
\begin{equation}
t_{23}=t_{23}^\mathrm{min}=\displaystyle \frac{8 \pi}{\beta}=\displaystyle \frac{16\pi }{\Omega_0^2} \mathcal{B}
\end{equation}
 Therefore, the efficiency  as a function of the bandwidth $ \mathcal{B}$ is:
\begin{equation}
\eta=\eta_0 \exp\left({\displaystyle -\frac{64\pi \mathcal{B}}{\Omega_0^2 T_2}}\right)
\label{eq:eff_B}
\end{equation}

We can also evaluate the number of pulses that can be stored at the same time while satisfying these conditions. In such a case, the signal is  composed of a pulse train. Each pulse has a during of $2 \pi/\mathcal{B}$, and the train has to fit within $t_{12}$. Therefore, the number of pulses is 
\begin{equation}
\label{nb_pulses}
t_{12} \times \frac{\mathcal{B}}{2\pi}=\displaystyle 4 \frac{ \mathcal{B}^2}{\Omega_0^2}
\end{equation} 
As we will see in section \ref{scaling}, this quantity should {\bf not be confused with the multiplexing capacity } because the efficiency is not constant as given by eq.\eqref{eq:eff_B}. 

\section{Experimental efficiency, storage time and bandwidth}
In the preceding section, we pointed out the interconnection between the ROSE protocol parameters imposed by realistic experimental conditions. In this section, we present an investigation of these constraints to evaluate the storage time and the bandwidth of the protocol. First, we investigated keeping the bandwidth constant  and increasing $t_{23}$ from its minimum value $t_{23}^\mathrm{min}$ corresponding to eq.(\ref{eq:eff}). Next, we varied the bandwidth and kept $t_{23}=t_{23}^\mathrm{min}$  corresponding to eq.(\ref{eq:eff_B}).
\subsection{Experimental setup}
The crystal under study, \yso, is a monoclinic crystal of space group C$_{2h}^6$ with a C$_2$ axis along the \baxis direction of the unit cell. Erbium substitutes at the two $Y$ sites, which have C$_1$ symmetry, and the behaviour of erbium in these sites has been extensively characterised by B\"ottger, Sun and coworkers \cite{Bottger06, Bottger06b, Bottger08, Bottger09}. Both  sites have transitions  from the ground state $^{4}$I$_{15/2}$ to the excited state $^{4}$I$_{13/2}$. We use ``site 1'' as described in \cite{Bottger09}, which is centered at 1536.48 nm. The crystal has three perpendicular optical extinction axes labelled \dun, \ddeux and \baxis, (the unit cell \baxis direction), which serve as a useful basis for the coordinate system. 

The 3 $\times$ 4   $\times$ 5 mm$^3$  sample, doped with 50 ppm of erbium, was cooled down in a variable temperature liquid helium cryostat to 1.8 K. Its optical depth was $\alpha L = 3.4$ along the $L=5$mm propagation dimension (\baxis axis). We applied a 2~T magnetic field $\vec{B}$ in the plane (\dun-\ddeux). $\vec{B}$  was aligned 135$^\circ$ anticlockwise with respect to \dun, which optimizes the coherence time as demonstrated by  B\"ottger \cite{Bottger09}.

A Koheras fiber laser was tuned on resonance with the transition between the $|-\frac{1}{2}\rangle$ Zeeman sublevels of the ground and excited states. After amplification, the output was split into two beams, one of which was used as the signal ($\sim$ 10 $\mu$W, see fig.\ref{fig:Rose_scheme}) while the other was used for the rephasing pulses ($\sim$~10~mW). Temporal and frequency  shaping of these beams was performed by acousto-optic modulators controlled with an arbitrary waveform generator (Tektronix AWG520). The signal and rephasing beam waists were  50 $\mu$m and 110 $\mu$m  respectively. They were polarized along \ddeux and \dun respectively \cite{Dajczgewand14}. The ROSE echo was recorded on an avalanche photodiode.

We estimated the Rabi frequency of the rephasing beam from an optical nutation experiment to be $\Omega_0=2\pi\times800$kHz.
As a consequence of this Rabi frequency and the constraints in Eqs. (\ref{adiabcond_exp}) and (\ref{t23_beta}), the minimum storage time was $ t_{23}^\mathrm{min}=2 t_{12}^\mathrm{min}=10\mu s$, occurring when $\mu=1$ and $\beta=2\pi\times400$kHz. For the different series of data, the signal is a single Gaussian pulse of duration $1.4\mu$s, which matches the minimum bandwidth. It is kept constant to simplify the characterization. For these minimum storage conditions, we obtained an efficiency of $24\%$.

\subsection{Efficiency as a function of the storage time}

\begin{figure}[h]
\centering{\includegraphics[width=0.6\paperwidth]{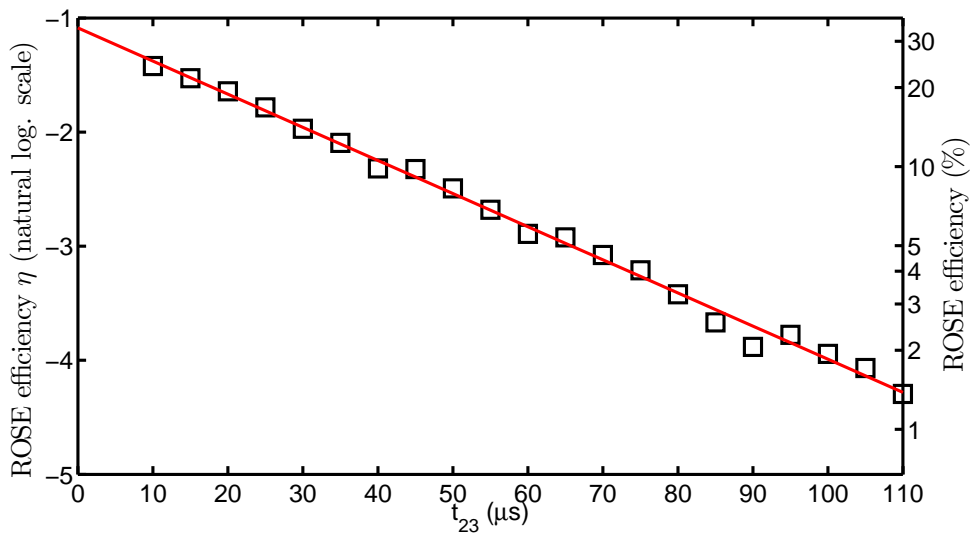}}
\caption{Exponential decay of the ROSE echo as the storage time $2 t_{23}$ is increased (squares). A linear fit (red line) gives the coherence time $T_2=138 \mu s$ and the maximum efficiency $\eta_0=34\%$ (eq. \ref{eq:eff}).}
\label{fig:Trace_vs_t23_article}
\end{figure}

Starting from the minimum storage time given above, we measured the decay of the efficiency when the storage time was increased from $ t_{23}^\mathrm{min}=10\mu s$ (eq. \ref{eq:eff}). This measurement gives the coherence time ultimately limiting the storage process. The bandwidth was kept to its minimal value $\mathcal{B}=2\pi\times800$kHz. We observed a clear exponential decay (fig. \ref{fig:Trace_vs_t23_article}).

From the linear fit, we obtain $T_2=138\mu s$. The value is smaller than previously reported in this material \cite{Dajczgewand14, Bottger09}, which is due to the lower magnetic field (2T compared to  3T in those works). Because optimizing the efficiency was not the main subject of the present work, we preferred to work at a lower field as this represents more stable conditions for long run experiments. The reader can refer to \cite{Dajczgewand14} for a more detailed discussion.

The fit in fig.\ref{fig:Trace_vs_t23_article} also gives the maximum efficiency $\eta_0=34\%$. This value is an extrapolation from the fit at zero storage time $t_{23} \rightarrow 0$ or in the limit of no decoherence. {It is consistent with the expected $\eta_0=(\alpha L)^{2} \exp({-\alpha L})=39\%$ for $\alpha L = 3.4$.}

\subsection{Efficiency as a function of the bandwidth}\label{section:bandwidth}

We studied the effect of the protocol bandwidth on the efficiency by increasing the interval $\mathcal{B}=2 \mu \beta$  chirped by the CHS pulses. We satisfied the conditions (\ref{adiabcond_exp}) and (\ref{t23_beta}), thus the bandwidth dependency was given by eq.(\ref{eq:eff_B}). We expect an exponential decay of the efficiency as a function of $\mathcal{B}$ due to the relationship between the bandwidth and the storage time. The slope of this decay should be given by the previously obtained fitting parameters $\eta_0=34\%$ and $T_2=138\mu s$ (from fig. \ref{fig:Trace_vs_t23_article}).

\begin{figure}[h]
\centering{\includegraphics[width=0.6\paperwidth]{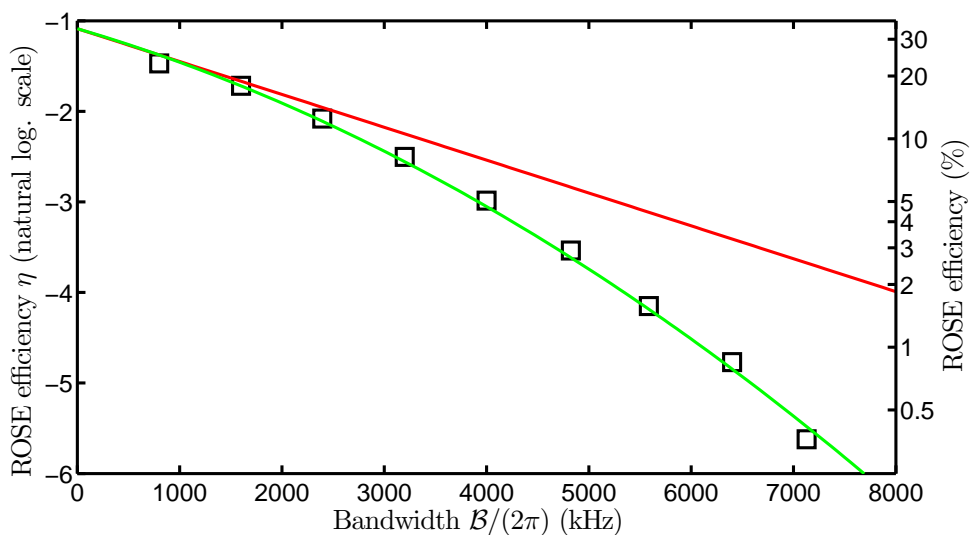}}
\caption{Efficiency as a function of the bandwidth $\mathcal{B}$ (squares). The red line corresponds to the expectation of eq.(\ref{eq:eff_B}) with the parameters $\eta_0=34\%$ and $T_2=138\mu s$ obtained from fig.\ref{fig:Trace_vs_t23_article}. The quadratic fit (green line) includes the influence of instantaneous spectral diffusion (see section \ref{section:ISD} for details).}
\label{fig:Trace_vs_Mu_article} 
\end{figure}

When $\mathcal{B}$ was varied from its initial value $2\pi\times800$kHz to $2\pi\times7100$kHz, the observed decay was clearly non-linear, as shown in fig.\ref{fig:Trace_vs_Mu_article}.
In this figure, the experimental data significantly deviate from the linear red curve that is consistent with the previous measurements (fig.\ref{fig:Trace_vs_t23_article}), indicating that the protocol efficiency  is affected by a bandwidth-dependent term. We ascribe this effect to instantaneous spectral diffusion (ISD). This is studied in detail in the following section since it is clearly a limiting factor on the memory performance.

\section{Influence of instantaneous spectral diffusion}\label{section:ISD}
ISD is the sudden shift in the transition frequency of an ion due to some change in the environment, such as the optical excitation of a nearby ion. Originally seen in NMR experiments (e.g \cite{Herzog56}), ISD is very commonly observed in a wide range of rare earth dopants and crystalline hosts. To give some examples,  Tb:YLiF$_4$ \cite{Liu90}, Tm:Y$_2$Si$_2$O$_7$ \cite{Wang96}, Eu:Y$_2$SiO$_5$ \cite{Mitsunaga92} and Pr:LaF$_3$ \cite{Zhang93}. A number of different interactions have been considered as the cause of ISD \cite{Graf98}, but the most common source in rare earth materials is the excitation-induced frequency shifts (EFS) between dopant ions, which are caused by a static electric or magnetic interaction coupling two ions that differs between the ground and excited state. Electric interactions tend to dominate in non-Kramers ions in low symmetry sites, while magnetic interactions become important for non-Kramers ions in high symmetry sites or for Kramers ions. The main exception is  Pr$^{3+}$ materials, which can exhibit ISD caused by non-equilibrium phonons \cite{Graf98}. In the following, we concentrate on EFS processes.

In a photon echo experiment or echo-like protocol such as ROSE, ISD is observed as a dependence of the echo amplitude on the density of ions excited by the $\pi$-pulses of the sequence. Because these excited atoms are spatially randomly distributed in the crystal, the ions that form the echo experience an inhomogeneous broadening. Unlike the normal static broadening in the crystal, this broadening occurs partway through the echo sequence and cannot be rephased, leading to an attenuation of the echo. The same attenuation of the echo is observed if the excited ions are spectrally resolved from the ions that form the echo, clearly differentiating ISD from resonant energy transfer processes. This technique of using a spectrally resolved ``scrambler'' pulse to measure ISD, first suggested by Mitsunaga {\it et al.} \cite{Mitsunaga92} has proven very useful for quantifying ISD, and we use it in this paper to confirm measurements of the ISD made using the ROSE protocol.

To understand the effect of ISD on the efficiency of ROSE, we first introduce it as a phenomenological parameter decreasing the measured coherence time $T_2$. This parameter can be extracted from the measurements of fig.\ref{fig:Trace_vs_Mu_article}. We then present an independent measurement of the ISD parameter using the standard photon-echo scrambler technique of Mitsunaga {\it et al.} \cite{Mitsunaga92}.  Finally, we compare these results to that obtained from a simple theoretical model of the electric and magnetic dipole-dipole coupling between erbium ions.

\subsection{Efficiency model including the ISD}\label{section:model_ISD}
In this section, we phenomenologically  introduce a bandwidth-dependent term into the coherence time $T_2(\mathcal{B})$ to account for the excess dephasing observed in the experiment. This can be written in terms of the homogeneous linewidth $1/T_2(\mathcal{B})$ \cite{Mossberg89, Liu90, Graf98} as:
\begin{equation}
\frac{1}{T_2(\mathcal{B})} =  \frac{1}{T_2^0} + \kappa \frac{\mathcal{B}}{2\pi}
\label{eq:T2_ISD}
\end{equation}
where $\kappa$ is the ISD coefficient. We choose to define $\kappa$ in radians but we use the units s$^{-1}$.Hz$^{-1}$, because $\kappa$ relates the excited bandwidth ${\mathcal{B}}/{(2\pi)}$ and the inverse of the coherence time ${1}/{T_2(\mathcal{B})}$ which are measured experimentally in units Hz and s$^{-1}$ respectively.

The efficiency scaling eq.\eqref{eq:eff_B} is modified accordingly:
\begin{equation}
\ln \left( \eta \right) =\ln \left( \eta_0 \right)  -\frac{64\pi }{\Omega_0^2 }\mathcal{B}\left( \frac{1}{T_2^0} + \kappa \frac{\mathcal{B}}{2\pi} \right)
\label{eq:eff_B_ISD}
\end{equation}
We retrieve the quadratic behavior observed experimentally (fig.\ref{fig:Trace_vs_Mu_article}). We can re-write this formula using the fitting parameters from fig.\ref{fig:Trace_vs_t23_article}:
\begin{equation}
\ln \left( \eta \right) =\ln \left( \eta_0 \right)  -\frac{64\pi }{\Omega_0^2 }\mathcal{B}\left( \frac{1}{T_2(\mathcal{B}_{800})} + \kappa \left[\frac{\mathcal{B}}{2\pi} - \frac{B_{800}}{2\pi}\right] \right)
\label{eq:eff_B_ISD}
\end{equation}
 with $\eta_0=34\%$, $T_2=138\mu$s, and $\mathcal{B}_{800}=2\pi\times800$kHz. The coefficient $ \kappa$ becomes the only free parameter in the fit. After optimization (least-square fitting), we obtain the green curve in fig.\ref{fig:Trace_vs_Mu_article}, for which $ \kappa=0.8$ s$^{-1}$.kHz$^{-1}$. 
From the fit we can determine the coherence time at the lowest excitation bandwidth: $T_2^0=\displaystyle \left({1}/{T_2(\mathcal{B}_{800})} -\kappa \times 800\mathrm{kHz} \right)^{-1}=151\mu$s. We also give a  few values of the modified coherence time as a function of the bandwidth to show the importance of the effect:

\begin{center}
\begin{tabular}{|l|c|c|c|c|}
  \hline
  $\mathcal{B}/(2 \pi)$ & 0 & 800 kHz & 5 MHz& 10 MHz\\   \hline
  $T_2(\mathcal{B})$ & 151$\mu$s & 138$\mu$s  & 94$\mu$s & 68$\mu$s\\
  \hline
\end{tabular}
\end{center}

\subsection{Independent measurement of the ISD coefficient}\label{indep_isd}
The ROSE protocol provides an interesting method to observe excitation induced dephasing as compared to the classic technique, used by Mitsunaga  {\it et al.} \cite{Mitsunaga92} and Graf  {\it et al.} \cite{Graf98}, for example. In this latter case, the traditional two-pulse echo is used to measure the coherence time of a target subgroup of ions and an independent laser beam (with a slightly different frequency named the ``scrambler'') is used to excite a second subgroup within the inhomogeneous profile. The two subgroups are then optically independent. If the ``scrambled" subgroup affects the coherence time of the target group, this means that the ions are magnetically or electrically coupled in some way. In the case of ROSE, the two subgroups overlap optically, which can make analysing the ISD more complex. For this reason, we decided to perform a spectroscopic measurement using the technique of Mitsunaga.

\begin{figure}[h]
\centering{\includegraphics[width=0.6\paperwidth]{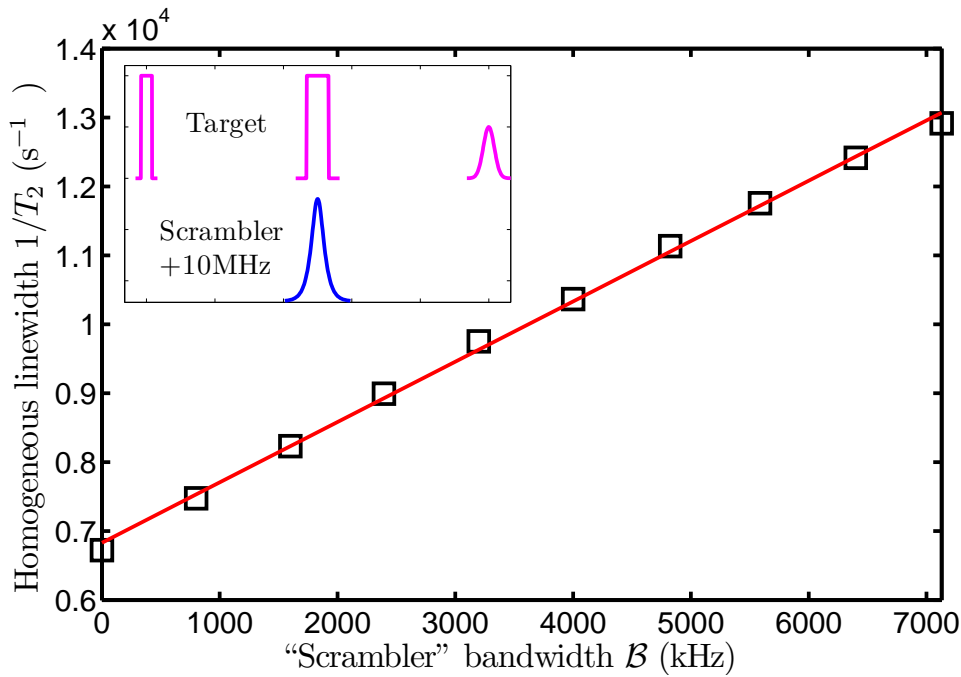}}
\caption{Homogeneous linewidth $\displaystyle \frac{1}{T_2}$ as a function of the ``scrambler" excited bandwidth.}
\label{fig:TraitSeries_2PE_article}
\end{figure}

We used a two-beam setup since it was already available from the ROSE setup. One beam excited the target ions whose coherence time $T_2$ was measured by a standard two-pulse echo sequence. The first pulse at $t=0$ was followed by a second one at $t_{12}$ (see inset of fig.\ref{fig:TraitSeries_2PE_article}). The beam was relatively weak, with a Rabi frequency of few hundreds of kHz corresponding to the target ions excited bandwidth. We fitted the decay of the echo as a function of  $t_{12}$ with an exponential decay, giving the coherence time of the target ions.

A stronger beam was shifted from the echo beam by 10MHz (AOM frequency difference). This scrambler beam excited a well separated spectral domain during the target ion echo sequence. To fully exploit our ROSE ``toolbox'', we used a CHS pulse for the scrambler which was synchronised at $t_{12}$ (see inset of fig. \ref{fig:TraitSeries_2PE_article}). Using a CHS pulse provides a reliable top-hat inversion profile \cite{de_Seze05}. Therefore, the definition of the excited bandwidth is relatively unambiguous:  $2 \mu \beta$ as defined in section \ref{section:scaling}. We varied the parameters of the CHS  while satisfying the condition (\ref{adiabcond_exp}) to ensure a constant inversion profile for the scrambled group. The scrambler bandwidth was varied from  $2\pi\times0$kHz (scrambler off) to $2\pi\times7100$kHz.

We can clearly observe an excitation induced dephasing of the target group by the scrambler beam in fig.\ref{fig:TraitSeries_2PE_article}. This dephasing is linear, as expected from eq.\eqref{eq:T2_ISD}, which supports our analysis in \ref{section:model_ISD}.

The linear fit gives an independent measurement of the ISD coefficient. We find $ \kappa_\mathrm{ind}=0.88$ s$^{-1}$.kHz$^{-1}$, close to the value of $ \kappa=0.8$ s$^{-1}$.kHz$^{-1}$ obtained in section \ref{section:model_ISD}. The agreement is close and will be discussed later on (see \ref{section:ISD_discussion}). 

The ISD coefficient has previously been measured in 0.02\% \eryso to be 1.3 $\times 10^{-12}$ Hz per excited ion per cm$^3$ \cite{Thiel12}. Those measurements were performed at an excitation density an order of magnitude higher than used here, but the values for the ISD  coefficient are in good agreement:  in the same units as the literature value, the ISD coefficient measured here by the ROSE technique is 1.1 $\times 10^{-12}$ Hz/(ion/cm$^3$), while it is 1.2 $\times 10^{-12}$ Hz/(ion/cm$^3$) for the two-pulse echo technique.

In the following section, we evaluate the dephasing induced by the static  electric and magnetic dipole coupling to determine if this can explain the observed ISD. This microscopic analysis is made possible by the previous measurements of the magnetic $g$-tensors by Sun {\it et al.} \cite{Bottger08}.

\subsection{Estimation of the ISD from microscopic parameters}\label{microscopic}
ISD typically arises from the inhomogeneous broadening caused by static electric or magnetic dipole-dipole interactions between the randomly distributed ions excited by a laser pulse. The amount of ISD is dependent on the difference in magnetic or electric dipole moments in the ground and excited states, and can be estimated using the inhomogeneous broadening model of Stoneham \cite{Stoneham69, Mims65}.  This model has been specifically applied to the situation of ISD  by Mims \cite{Mims65}, and we only briefly describe it here. In Stoneham's model, the crystal is treated as uniform and isotropic, reasonable assumptions for the low excitation densities used in our experiments. Under these assumptions, the FWHM inhomogeneous broadening caused by any dipolar interaction is
 \begin{equation}
 \Delta \omega = \frac{16 \pi^2}{9\sqrt{3}}A n_e
 \end{equation}
 where $A$ is a constant describing the interaction and $n_e$ is the spatial density of excited ions. One of the problems of applying this equation to optical ISD is that the spatial density can be difficult to estimate in a photon echo experiment,  as excitation density is spectrally and spatially inhomogeneous when simple square or Gaussian pulses are used. A new ISD measurement method recently described by Thiel {\it et al.}  \cite{Thiel14} gets around this problem, allowing $n_e$ to be estimated directly from the ISD data. 
 
 This method is not necessary in our experiment because the complex hyperbolic shape of the exciting pulses means that they reliably invert every ion in the excited bandwidth, and so $n_{e}$ can be easily calculated. For  the Lorentzian shape of the inhomogeneous absorption line and assuming the ions are excited  at the line center, $n_e$ is  
 \begin{equation}
 n_e = n_Y C \frac{2}{\pi \Gamma_\mathrm{inh}}\mathcal{B}
 \end{equation}
 where $n_Y$ is the spatial density of yttrium ions in the structure, $C$ the dopant concentration and
 $ \Gamma_\mathrm{inh}$ the inhomogeneous line width.  In \yso, $n_Y=1.83 \times 10^{22}$ cm$^{-3}$, and as erbium substitutes equally at the two yttrium sites in \yso, $C=\frac{1}{2}\times 50$ ppm, while $\Gamma_\mathrm{inh}=2\pi\times 630$MHz.  In materials where multiple ground states are populated, $C$ needs to take into account the occupancy of the resonant level, but for \eryso below 3~K the population of the upper ground state is negligible.
 
 The excitation-induced broadening causes a modification of the coherence time measured by the two-pulse echo:
 \begin{equation}
 \frac{1}{T_2(\mathcal{B})} = \frac{1}{T_2^0}+\frac{1}{4}\Delta \omega
 \end{equation}
which gives an ISD coefficient:
\begin{equation}
\kappa_\mathrm{\mu} = \frac{8 \pi^3}{9\sqrt{3}}A n_Y C \frac{2}{\pi \Gamma_\mathrm{inh}}\label{kappa_mu}
\end{equation} 
 
The constant $A$ can have contributions from both a magnetic dipole-dipole interaction and an electric dipole-dipole interaction. For the former,
\begin{equation}
 A_{mag} = \frac{\mu_0 \hbar}{4\pi}|\Delta\vec{\mu}_{mag}|^2
 \label{eq:mdd}
 \end{equation}
 In this equation, $\Delta\vec{\mu}_{mag}$ is the difference in expectation values of the magnetic moment (either electronic or nuclear) in the ground and excited states, and has units of  T$^{-1}$.rad.s$^{-1}$. For \eryso, the relevant moment is the electronic dipole moment, and the coupling constant can be accurately calculated from the published moment tensor \cite{Bottger08} to be $A_{mag} = 2.8 \times 10^{-19}$ m$^3$.rad.s$^{-1}$ for the transition and magnetic field of interest. For an electric dipole-dipole interaction,
  \begin{equation}
 A _{el} = \frac{1}{4\pi \epsilon_r \epsilon_0 \hbar}|\Delta\vec{\mu}_{el}|^2
 \label{eq:edd}
 \end{equation}  
 for $\Delta\vec{\mu}_{el}$ in units of  C.m. It is much more difficult to estimate the value of this parameter as, to the authors' knowledge, the Stark shift of \eryso has not been published. However, the Stark shift does not vary  much between different rare earth ions and hosts,  typically lying between $10 - 100$ kHz/(V.cm$^{-1}$) \cite{Macfarlane07}. To get an estimation for the electric dipole moment, we take the value of the Stark shift for Er:LiNbO$_3$, 25 kHz/(V.cm$^{-1}$) \cite{Hastings-Simon06}, which corresponds to $\Delta\vec{\mu}_{el} = 1.65 \times 10^{-31}$ C.m,  as this is the only published value we have found for the $^{4}$I$_{15/2} - ^{4}$I$_{13/2}$  transition.
Using the relative permittivity  perpendicular to the C$_2$ axis $\epsilon_r = 4$ \cite{Hedges11}, $A_{el}= 5.8\times 10^{-19}$ m$^3$.rad.s$^{-1}$, slightly larger than $A_{mag}$. This value is only valid to within an order of magnitude, first, because the actual Stark shift is unknown and second, because the relative permittivity is very anisotropic in \yso.
 
 Taking the values of $A_{mag}$ and $A_{el}$ above, we calculate that  $\kappa_{\mu} = 0.33$ s$^{-1}$.kHz$^{-1}$ if only the magnetic dipole-dipole interaction is considered, while if we consider both magnetic and electric interactions,  $\kappa_{\mu} = 1$ s$^{-1}$.kHz$^{-1}$.

\subsection{Discussion}\label{section:ISD_discussion}

The theoretical prediction for the ISD coefficient based on the assumption of a combination of magnetic and electric dipole-dipole interactions, $\kappa_{\mu} =1$ s$^{-1}$.kHz$^{-1}$, is in very good agreement with the experimental values of $ \kappa=0.8$ s$^{-1}$.kHz$^{-1}$ and $\kappa_\mathrm{ind}=0.88$ s$^{-1}$.kHz$^{-1}$ obtained in \ref{section:model_ISD} and \ref{indep_isd} respectively, given that the electric dipole contribution could only be approximated. This agreement confirms that a mixture of electric and magnetic dipole-dipole interactions dominates the ISD. A more accurate estimation of the electric dipole-dipole contribution would require a measurement of the Stark shift  as well as the dielectric constant tensor, which is beyond the scope of this paper. 

The theoretical prediction showed that the magnetic dipole-dipole interaction is only half the size of the electric dipole-dipole interaction, despite the large electronic magnetic moment of Er$^{3+}$. This highlights the importance of having a difference in the moment between the optical ground and excited states. While the ground and excited state moments are $\mathcal{O}(100)$~GHz/T, the difference between them is  $\mathcal{O}(10)$~GHz/T for the lowest-to-lowest transition. This means that the optical magnetic dipole-dipole ISD is relatively low compared to the ISD that would be observed on the spin transition.

\section{ROSE performance including the instantaneous spectral diffusion}
ISD will be a critical limitation on the ability to extend the protocol bandwidth. We evaluate this limitation in this section by using the previous measurements to derive quantitative scaling laws for the performance.

\subsection{Storage time, efficiency and multiplexing capacity scaling}\label{scaling}

As discussed in section \ref{section:scaling}, even in the absence of ISD, the different figures of merit for the storage protocol are not independent. For example, the adiabatic condition eq.\eqref{adiabcond} constrains the obtainable efficiency and bandwidth eq.\eqref{eq:eff_B} for a fixed coherence time $T_2$ (without ISD) and laser power $\Omega_0^2$. The ISD will place further limits on the figures of merit. In this section, we analyze the influence of the ISD on the protocol performances --storage time, efficiency and multiplexing capacity -- and then apply this analysis to the specific case of erbium in \yso.

\begin{itemize}
\item Storage time: ISD directly limits the coherence time and thus the storage time, which scales as $\displaystyle \frac{2\pi}{\kappa \mathcal{B}}$ as soon as the ISD is sufficiently large to dominate the ``bare" coherence time $T_2^0$ (eq. \ref{eq:T2_ISD}).
\item Efficiency: ISD introduces a quadratic decay of the efficiency  as a function of the bandwidth (log scale, eq. \ref{eq:eff_B_ISD}). However, this decay can be compensated by a larger laser power $\Omega_0^2$. In that case, the time sequence must be shortened to avoid the decoherence at the price of a shorter storage time.
\item Multiplexing capacity: The temporal multiplexing capacity is the ability to address different spectral channels in the inhomogeneous linewidth, or equivalently to store a series of pulses in parallel. A high potential multiplexing capacity is a major advantage of the ROSE protocol.  The temporal multiplexing capacity is equivalent to the time-to-bandwidth product $\displaystyle T_2(\mathcal{B}) \times \frac{\mathcal{B}}{2\pi}$ where $T_2(\mathcal{B}) $ is the inverse of the spectral resolution. This general argument can be applied to our specific protocol. As we discuss for eq.\eqref{nb_pulses}, the number of temporal modes and the bandwidth can always be increased to the detriment of the efficiency. The multiplexing capacity has a significant meaning only when the efficiency is constant. In practice, defining the capacity as $\displaystyle T_2(\mathcal{B}) \times {\mathcal{B}}/{(2\pi)}$ means that we keep $t_{12}= T_2(\mathcal{B})$ (see eq.\ref{nb_pulses}) and the storage time  $2t_{23}= 4T_2(\mathcal{B})$. This gives a rather weak criterium for the efficiency which is only $\eta= \eta_0 \exp\left(-8\right)$ (eq.\ref{eq:eff}). A different criterium can be derived from the previous expressions.

In our case and in the presence of ISD, the capacity defined as $\displaystyle T_2(\mathcal{B}) \times {\mathcal{B}}/{(2\pi)}$ is no longer limited by the inhomogeneous linewidth  but  instead scales as $1/\kappa$.
\end{itemize}

Clearly, ISD limits the memory performance, particularly when high bandwidth storage is considered. It is, therefore, important to try to minimise the effect of ISD. As Eq. \ref{kappa_mu} shows, there are two main ways to do this: reduce the erbium concentration $C$, or reduce the dipole-dipole coupling constants $A_{el}$ and $A_{mag}$. Reducing the concentration is a simple way of reducing the ISD, as the ISD scales linearly with the concentration. Because the optical density also decreases with the concentration, it is necessary to increase the length of the crystal to maintain the same total absorption $\alpha L$ and therefore the same protocol efficiency. 

The other option to reduce ISD is to reduce the dipole coupling constants by minimising the difference in ground and excited state dipole moments. For the electric dipole-dipole interaction, this is not possible as the Stark shift for any given electronic transition is fixed. However, the difference in nuclear magnetic dipole moments is strongly dependent on the direction of the applied field. In the following section we investigate whether the memory performance can be improved by rotating the crystal within the magnetic field.

\subsection{Optimization strategy in \eryso}

As demonstrated by B\"ottger and coworkers \cite{Bottger09}, the coherence properties of \eryso strongly depend on the orientation of the magnetic field, so there is the possibility that the ISD can be reduced by rotating the field. In what follows, we consider only fields in the plane (\dun,\ddeux). Out of this plane, the two magnetically nonequivalent subsites are split in the optical spectrum, reducing the optical depth and the protocol efficiency as a consequence. Our choice also simplifies the analysis.

\begin{figure}[h]
\centering{\includegraphics[width=0.7\paperwidth]{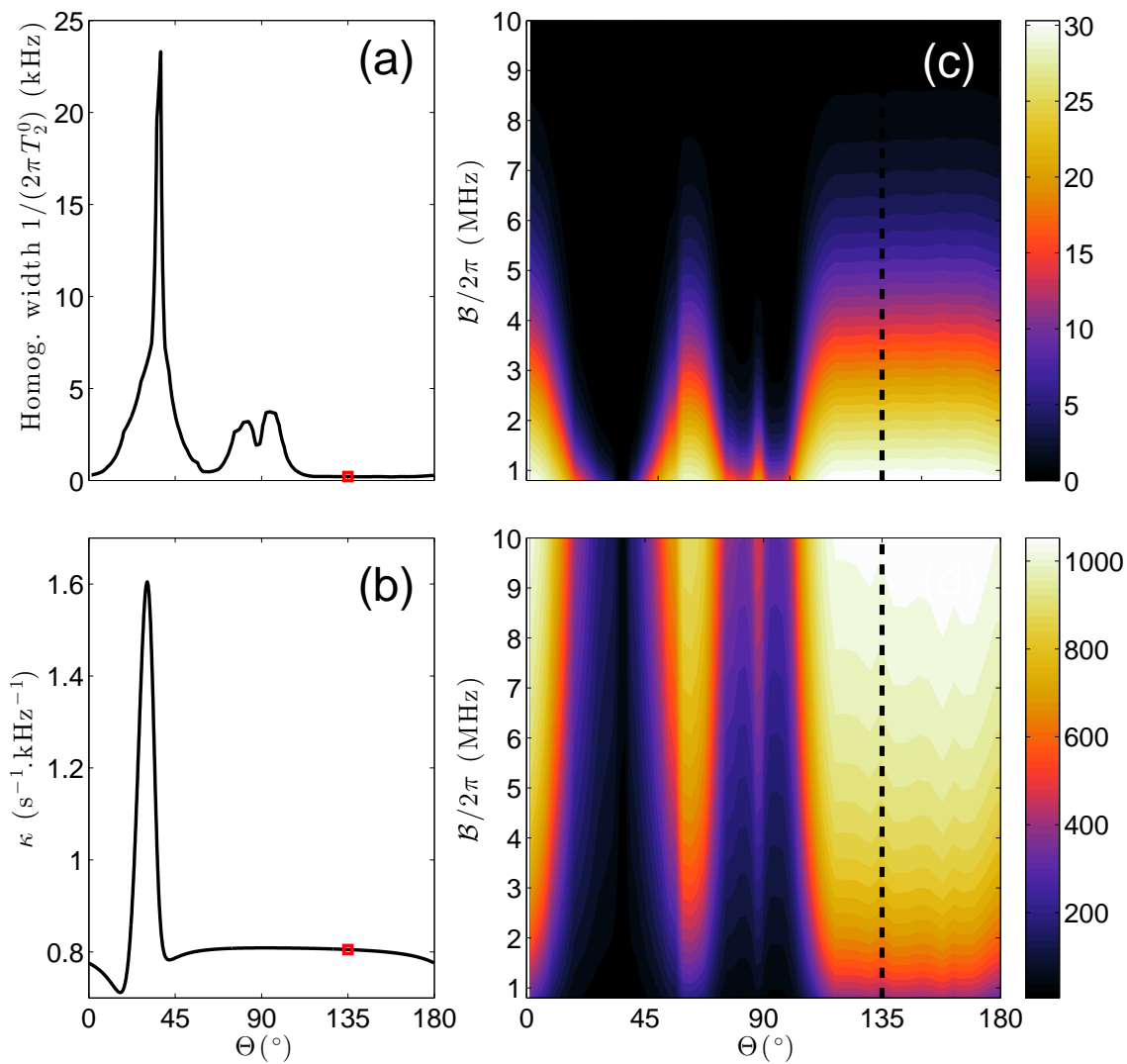}}
\caption{(a) (From \cite{Bottger09}) Homogeneous linewidth $\displaystyle \frac{1}{2 \pi T_2^0}$ at 3 T for 15 ppm  concentration as a function of $\Theta$, the magnetic field orientation in the plane (\dun,\ddeux). (b) Theoretical variation in $\kappa$ as a function of $\Theta$. (c) Estimated ROSE protocol efficiency, in percent, as a function of the bandwidth $\mathcal{B}$ and the orientation $\Theta$. (d) Time-to-bandwidth product as a function of the bandwidth $\mathcal{B}$ and the orientation $\Theta$. In all figures, the orientations $\Theta= 0$ or $180^\circ$ and $\Theta= 90^\circ$ correspond to \dun and \ddeux respectively. The two red squares in (a) and (b) and the dashed lines in (c) and (d) correspond to the experimental condition $\Theta= 135^\circ$ of figs. \ref{fig:Trace_vs_Mu_article} and \ref{fig:TraitSeries_2PE_article}.}
\label{fig:map_phi}
\end{figure}

To properly evaluate the performance of the memory as a function of the magnetic field direction, the dependence of $T_2^0$  and $\kappa$ on the field should be known. However, measuring these two parameters as function of the magnetic field rotation angle $\Theta$ is experimentally challenging. Therefore, we use a previous measurement of  $T_2^0(\Theta)$   \cite{Bottger09} (see fig. \ref{fig:map_phi}.a) which was performed at a lower concentration (15ppm). Meanwhile, for  $\kappa$, we start with our measured value $ \kappa=0.8$ s$^{-1}$.kHz$^{-1}$, which corresponds to an angle from the \dun axis of $\Theta= 135^\circ$,  and assume that the magnetic dipole-dipole contribution to this number is as given in Section \ref{microscopic}, 0.33 s$^{-1}$.kHz$^{-1}$, while the electric dipole accounts for the remainder. The change in $\kappa$ with magnetic field orientation due to the change in the magnetic dipole coupling constant is then given by eq. \ref{eq:mdd}.

The angular dependence of $T_2^0(\Theta)$  and $\kappa(\Theta)$ is shown in figs.\ref{fig:map_phi}.a and \ref{fig:map_phi}.b. The resemblance between the two curves is clear. This is expected, because the larger $\Delta\vec{\mu}(\Theta)$ is, the more sensitive the transition frequency to any fluctuating local magnetic field is (spin flip-flops of neighboring erbium). Therefore, when $\kappa(\Theta)$ is large a larger decoherence rate is also expected. This rudimentary argument explains the existence of the peak for both $T_2^0(\Theta)$ and $\kappa(\Theta)$ close to $\Theta= 35^\circ$. The presence of lobes close to $90^\circ$ in fig.\ref{fig:map_phi}.a is discussed in \cite{Bottger09}.

Starting from the estimation above of $T_2^0(\Theta)$  and $\kappa(\Theta)$ for different orientations $\Theta$ of the magnetic field, we can compute the expected performance of the protocol in \eryso. Here, we estimate the efficiency (eq.\ref{eq:eff_B_ISD}) and the time-to-bandwidth product $\displaystyle T_2(\mathcal{B}) \times \frac{\mathcal{B}}{2\pi}$ (from eq.\ref{eq:T2_ISD}) as function of the bandwidth $\mathcal{B}$ for the different orientations $\Theta$. We obtain 2-dimensional plots in fig.\ref{fig:map_phi}.c and \ref{fig:map_phi}.d respectively.

These plots clearly show that the orientation of the field along $\Theta= 35^\circ$ or $\Theta= 90^\circ$ is unfavorable and should be avoided, while the range $\Theta= 120^\circ-180^\circ$ is preferable. It justifies {\it a posteriori} our experimental choice of $\Theta= 135^\circ$. However, even at these fields we observe a substantial decay of the efficiency (fig.\ref{fig:map_phi}.c) as soon as the bandwidth is increased. As an example, for a bandwidth of $\mathcal{B}=2\pi\times10$MHz, the expected efficiency is only $0.30\%$ with $\Theta= 135^\circ$. This rapid decay is mainly explained by the ISD. In the absence of ISD, the efficiency would be $17\%$ for the same coherence time. Concerning the time-to-bandwidth product, this rapidly saturates to $1/\kappa \sim 1200$ with $\kappa=0.8$s$^{-1}$.kHz$^{-1}$.

Overall, the best performance can be obtained with an orientation of the field between $\Theta= 120^\circ$ and $180^\circ$, while a bandwidth of $2-4$MHz seems to be a good trade-off between an efficiency of $\sim 20$\% and a time-to-bandwidth product of $\sim 800$. It should be noted that theses figures would be obtained for a coherence time of $\sim 800\mu$s as demonstrated in  \cite{Bottger09} (corresponding to fig.\ref{fig:map_phi}.c).

\subsection{Effect of ISD in other memory protocols}
Most of the quantum memory schemes proposed for rare earth ions are photon echo based, and we have seen that ISD is a general consequence of the $\pi$-pulses applied in a photon echo sequence. It can be expected, then, that it will have an effect on other memory protocols. This section contains a very general discussion of the effects of ISD in different memory protocols. We will compare  three protocols: spectral engraving-based protocols (AFC, GEM), non-engraving based protocols (ROSE, HYPER, RASE), and the case of spin storage where dynamic decoupling sequences are used to lengthen storage time.

The protocols that do not involve engraving all operate in a similar way to ROSE: the bandwidth or multimode capacity is given by the bandwidth of the $\pi$-pulses used, and these pulses can invert every ion in the inhomogeneous line that is within the bandwidth. This can be a large proportion of the ions in the crystal because the memories are typically operated near the center of the inhomogeneous profile in order to have a high optical depth and therefore high efficiency. Therefore, non-engraving type memories can be expected to show significant ISD. To avoid the ISD, it is necessary to obtain the high optical density needed without having a high excitation density. One solution to this problem is to place the memory in a resonant optical cavity, as has been suggested for RASE \cite{beavan12}.

Engraving based memory protocols can also involve optical $\pi$-pulses with a bandwidth given by the protocol bandwidth. However, the first step of such a memory protocol involves spectrally tailoring the protocol bandwidth to create a grating (AFC) or a narrow feature (GEM). During this process, the majority of ions within the bandwidth are optically pumped to non-resonant levels, so that the protocol is enacted on a very low density of ions. For this reason, ISD has less effect on the storage step of an engraving-based memory. However, ISD can be expected to have some effect on the engraving step of the memory protocol. When ions are excited in the hole burning process, they shift the transition frequencies of nearby ions, which will shift some non-resonant ions into resonance with the laser, allowing them to be optically pumped. The width of a spectral hole, or structure, burnt in the presence of ISD will be wider than one burnt in the absence of ISD. This can affect the memory performance as the engraving-based memories require sharp spectral features. This effect can be minimized by using slow pumping rates.

The final case we consider is the effect of dynamic decoupling sequences applied to electron spin or hyperfine transitions. In a typical dynamic decoupling sequence, a series of  $\pi$-pulses is applied to the spin transition at a particular rate. The effect of any spectral diffusion processes slower than the repetition rate is removed by the pulse sequence, resulting in a longer storage time for the system. The $\pi$-pulses typically cover the entire spin linewidth, and so can change the energy level of every rare earth ion in the crystal. This leads to instantaneous spectral diffusion proportional to the difference in magnetic moments between the two levels.  For electron spin transitions this difference is large. Even for hyperfine transitions, which have much smaller magnetic moments, the high density of excited ions and the large number of pulses means that a substantial amount of ISD is possible. As the effect is cumulative, it will limit the total number of pulses that can be applied without degrading the coherence. One method of avoiding this ISD is to choose magnetic field directions along which the magnetic moment difference is minimized (ZEFOZ points) \cite{fraval04}.

Finally, we note that in this section, we have only considered how ISD will affect the different memory protocols in general. The amount of ISD seen in a particular memory will be highly dependent on the rare earth ion and host material, and the particular parameters of the protocol used experimentally.
 
\section{Conclusion}
We studied the dependence of the storage efficiency of the ROSE protocol in 
\eryso as a function of the storage bandwidth. We observed a decrease in the efficiency with bandwidth which we attributed to instantaneous spectral diffusion. This effect is particularly important for the ROSE protocol because it involves strong optical excitation, and will affect the ability to extend the bandwidth of the memory protocol. As the ISD is proportional to the dopant concentration, a simple first step to reduce this effect will be to reduce the dopant concentration.

Approximately one third of the ISD was attributed to a magnetic dipole-dipole interaction, and it was shown that this contribution can be minimised by an appropriate choice of the magnetic field direction in the (\dun,\ddeux) plane. A more detailed study of the magnetic interactions away from this plane, and for erbium in other host matrices, would certainly be valuable.

The remainder of the observed ISD was attributed to an electric dipole-dipole interaction, which scales with the Stark shift of the electronic transition. As the Stark shifts for different rare earth ions and in different hosts do not vary substantially, similar levels of ISD can be expected in other memory materials. The effect of ISD on the performance of different memories will, therefore, be largely dependent  on the optical excitation density during storage process.

\section*{Acknowledgments}

This work is supported by the French national grant RAMACO no. ANR-12-BS08-0015-02. The research leading to these results has received funding from the People Programme (Marie Curie Actions) of the European Union's Seventh Framework Programme FP7/2007-2013/ under REA grant agreement no. 287252.

\section*{References}
%\bibliographystyle{iopart-num}
%\bibliography{BW_ROSE_ISD_bib,ROSE_efficiency_bib}
\bibliographystyle{iopart-num}
\bibliography{bib_BW_ROSE_ISD}

\providecommand{\newblock}{}
\begin{thebibliography}{10}
\expandafter\ifx\csname url\endcsname\relax
  \def\url#1{{\tt #1}}\fi
\expandafter\ifx\csname urlprefix\endcsname\relax\def\urlprefix{URL }\fi
\providecommand{\eprint}[2][]{\url{#2}}

\bibitem{Sangouard11}
Sangouard N, Simon C, de~Riedmatten H and Gisin N 2011 {\em Rev. Mod. Phys.\/}
  {\bf 83}(1) 33--80

\bibitem{Collins07}
Collins O~A, Jenkins S~D, Kuzmich A and Kennedy T~A~B 2007 {\em Phys. Rev.
  Lett.\/} {\bf 98}(6) 060502

\bibitem{Simon07}
Simon C, de~Riedmatten H, Afzelius M, Sangouard N, Zbinden H and Gisin N 2007
  {\em Phys. Rev. Lett.\/} {\bf 98} 190503

\bibitem{Simon10}
Simon C, de~Riedmatten H and Afzelius M 2010 {\em Phys. Rev. A\/} {\bf 82}(1)
  010304

\bibitem{Lan09}
Lan S~Y, Radnaev A~G, Collins O~A, Matsukevich D~N, Kennedy T~A and Kuzmich A
  2009 {\em Opt. Express\/} {\bf 17} 13639--13645

\bibitem{Golubeva11}
Golubeva T, Golubev Y, Mishina O, Bramati A, Laurat J and Giacobino E 2011 {\em
  Phys. Rev. A\/} {\bf 83}(5) 053810

\bibitem{Nunn08}
Nunn J, Reim K, Lee K~C, Lorenz V~O, Sussman B~J, Walmsley I~A and Jaksch D
  2008 {\em Phys. Rev. Lett.\/} {\bf 101} 260502

\bibitem{Afzelius09}
Afzelius M, Simon C, de~Riedmatten H and Gisin N 2009 {\em Phys. Rev. A\/} {\bf
  79} 052329

\bibitem{Bonarota11}
Bonarota M, Le~Gou{\"e}t J~L and Chaneli{\`e}re T 2011 {\em New J. Phys.\/}
  {\bf 13} 013013

\bibitem{sinclair14}
Sinclair N, Saglamyurek E, Mallahzadeh H, Slater J~A, George M, Ricken R,
  Hedges M~P, Oblak D, Simon C, Sohler W and Tittel W 2014 {\em Phys. Rev.
  Lett.\/} {\bf 113}(5) 053603

\bibitem{Tittel10}
Tittel W, Afzelius M, Cone R, Chaneli{\`e}re T, Kr{\"o}ll S, Moiseev S and
  Sellars M 2010 {\em Laser Photon. Rev.\/} {\bf 4} 244--267

\bibitem{Lvovsky09}
Lvovsky A~I, Sanders B~C and Tittel W 2009 {\em Nature Photon.\/} {\bf 3}
  706--714

\bibitem{Ruggiero09}
Ruggiero J, {Le Gou\"{e}t} J~L, Simon C and Chaneli\`{e}re T 2009 {\em Phys.
  Rev. A\/} {\bf 79} 053851

\bibitem{Sangouard10}
Sangouard N, Simon C, Min\'a\v{r} J, Afzelius M, Chaneli\`ere T, Gisin N,
  Le~Gou\"et J~L, de~Riedmatten H and Tittel W 2010 {\em Phys. Rev. A\/} {\bf
  81}(6) 062333

\bibitem{Nilsson05}
Nilsson M and Kr\"oll S 2005 {\em Opt. Commun.\/} {\bf 247} 393--403

\bibitem{Moiseev01}
Moiseev S~A and Kr\"oll S 2001 {\em Phys. Rev. Lett.\/} {\bf 87} 173601

\bibitem{Alexander06}
Alexander A~L, Longdell J~J, Sellars M~J and Manson N~B 2006 {\em Phys. Rev.
  Lett.\/} {\bf 96}(4) 043602

\bibitem{Kraus06}
Kraus B, Tittel W, Gisin N, Nilsson M, Kr\"oll S and Cirac J~I 2006 {\em Phys.
  Rev. A\/} {\bf 73}(2) 020302

\bibitem{Damon11}
Damon V, Bonarota M, Louchet-Chauvet A, Chaneli\`ere T and {Le~Gou\"et} J~L
  2011 {\em New J. Phys.\/} {\bf 13} 093031

\bibitem{Hyper}
McAuslan D~L, Ledingham P~M, Naylor W~R, Beavan S~E, Hedges M~P, Sellars M~J
  and Longdell J~J 2011 {\em Phys. Rev. A\/} {\bf 84}(2) 022309

\bibitem{Dajczgewand14}
Dajczgewand J, {Le Gou\"{e}t} J~L, Louchet-Chauvet A and Chaneli\`{e}re T 2014
  {\em Opt. Lett.\/} {\bf 39} 2711--2714

\bibitem{de_Seze05}
de~Seze F, Dahes F, Crozatier V, Lorger{\'e} I, Bretenaker F and {Le Gou{\"e}t}
  J~L 2005 {\em Eur. Phys. J. D\/} {\bf 33} 343

\bibitem{Pascual-Winter13}
Pascual-Winter M~F, Tongning R~C, Chaneli\`ere T and {Le Gou\"et} J~L 2013 {\em
  New J. Phys.\/} {\bf 15} 055024

\bibitem{Hioe84}
Hioe F~T 1984 {\em Phys. Rev. A\/} {\bf 30} 2100--2103

\bibitem{Silver85}
Silver M~S, Joseph R~I and Hoult D~I 1985 {\em Phys. Rev. A\/} {\bf 31}
  2753--2755

\bibitem{Bottger06}
B\"ottger T, Thiel C~W, Sun Y and Cone R~L 2006 {\em Phys. Rev. B\/} {\bf
  73}(7) 075101

\bibitem{Bottger06b}
B\"ottger T, Sun Y, Thiel C~W and Cone R~L 2006 {\em Phys. Rev. B\/} {\bf
  74}(7) 075107

\bibitem{Bottger08}
Sun Y, B\"ottger T, Thiel C~W and Cone R~L 2008 {\em Phys. Rev. B\/} {\bf
  77}(8) 085124

\bibitem{Bottger09}
B\"ottger T, Thiel C~W, Cone R~L and Sun Y 2009 {\em Phys. Rev. B\/} {\bf
  79}(11) 115104

\bibitem{Herzog56}
Herzog B and Hahn E~L 1956 {\em Phys. Rev.\/} {\bf 103} 148--166

\bibitem{Liu90}
Liu G~K and Cone R~L 1990 {\em Phys. Rev. B\/} {\bf 41}(10) 6193--6200

\bibitem{Wang96}
Wang G~M, Leask M~J~M, Godfrey K~W, Equall R~W, Cone R~L and Wondre F~R 1996
  {\em Opt. Lett.\/} {\bf 21} 818--820

\bibitem{Mitsunaga92}
Mitsunaga M, Takagahara T, Yano R and Uesugi N 1992 {\em Phys. Rev. Lett.\/}
  {\bf 68} 3216--3219

\bibitem{Zhang93}
Zhang J and Mossberg T~W 1993 {\em Phys. Rev. B\/} {\bf 48} 7668--7671

\bibitem{Graf98}
Graf F~R, Renn A, Zumofen G and Wild U~P 1998 {\em Phys. Rev. B\/} {\bf 58}(9)
  5462--5478

\bibitem{Mossberg89}
Huang J, Zhang J~M, Lezama A and Mossberg T~W 1989 {\em Phys. Rev. Lett.\/}
  {\bf 63}(1) 78--81

\bibitem{Thiel12}
Thiel C~W, Babbitt W~R and Cone R~L 2012 {\em Physical Review B\/} {\bf 85}
  174302

\bibitem{Stoneham69}
Stoneham A~M 1969 {\em Rev. Mod. Phys.\/} {\bf 41}(1) 82--108

\bibitem{Mims65}
Mims W~B 1965 Electron spin echoes {\em Electron Paramagnetic Resonance\/} ed
  Geschwind S (Plenum, New York, 1972) pp 263--351

\bibitem{Thiel14}
Thiel C~W, Macfarlane R~M, Sun Y, B\"ottger T, Sinclair N, Tittel W and Cone
  R~L 2014 {\em Laser Phys.\/} {\bf 24} 106002

\bibitem{Macfarlane07}
Macfarlane R~M 2007 {\em J. Lumin.\/} {\bf 125} 156--174

\bibitem{Hastings-Simon06}
Hastings-Simon S~R, Staudt M~U, Afzelius M, Baldi P, Jaccard D, Tittel W and
  Gisin N 2006 {\em Opt. Commun.\/} {\bf 266} 716--719

\bibitem{Hedges11}
Hedges M~P 2011 {\em High Performance Solid State Quantum Memory\/} {PhD}
  thesis The Australian National University

\bibitem{beavan12}
Beavan S~E, Hedges M~P and Sellars M~J 2012 {\em Phys. Rev. Lett.\/} {\bf 109}
  093603

\bibitem{fraval04}
Fraval E, Sellars M~J and Longdell J~J 2004 {\em Phys. Rev. Lett.\/} {\bf 92}
  077601

\end{thebibliography}

\end{document}